\documentclass[conference]{IEEEtran}
\IEEEoverridecommandlockouts
\usepackage{cite}
\usepackage{amsmath,amssymb,amsfonts}
\usepackage{algorithmic}
\usepackage{graphicx}
\usepackage{textcomp}
\usepackage{xcolor}
\usepackage{svg}
\usepackage{amsmath}
\usepackage{subcaption}

\usepackage[font=small]{caption}

\def\BibTeX{{\rm B\kern-.05em{\sc i\kern-.025em b}\kern-.08em
    T\kern-.1667em\lower.7ex\hbox{E}\kern-.125emX}}
\begin{document}

\title{\vspace{-4mm} Multi-Modal Transformer and Reinforcement Learning-based Beam Management\\ \vspace{-4mm}}

\author{\IEEEauthorblockN{Mohammad Ghassemi\textsuperscript{1}, Han Zhang\textsuperscript{1}, Ali Afana\textsuperscript{2}, Akram Bin Sediq\textsuperscript{2}, \\ and Melike Erol-Kantarci\textsuperscript{1}, \textit{Senior Member}, \textit{IEEE}}
\IEEEauthorblockA{\textit{\textsuperscript{1}School of Electrical Engineering and Computer Science, University of Ottawa, Ottawa, Canada} \\
\textit{\textsuperscript{2}Ericsson Inc., Ottawa, Canada}\\
Emails:\{mghas017, hzhan363, melike.erolkantarci\}@uottawa.ca,
\{ali.afana, akram.bin.sediq\}@ericsson.com}

\vspace{-10mm}}

\maketitle

\vspace{-2mm}
\begin{abstract}
Beam management is an important technique to improve signal strength and reduce interference in wireless communication systems. Recently, there has been increasing interest in using diverse sensing modalities for beam management. However, it remains a big challenge to process multi-modal data efficiently and extract useful information. 
On the other hand, the recently emerging multi-modal transformer (MMT) is a promising technique that can process multi-modal data by capturing long-range dependencies. While MMT is highly effective in handling multi-modal data and providing robust beam management, integrating reinforcement learning (RL) further enhances their adaptability in dynamic environments.
In this work, we propose a two-step beam management method by combining MMT with RL for dynamic beam index prediction.
In the first step, we divide available beam indices into several groups and leverage MMT to process diverse data modalities to predict the optimal beam group. In the second step, we employ RL for fast beam decision-making within each group, which in return maximizes throughput. Our proposed framework is tested on a 6G dataset. In this testing scenario, it achieves higher beam prediction accuracy and system throughput compared to both the MMT-only based method and the RL-only based method.
\end{abstract}

\vspace{-2mm}
\section{Introduction}
\vspace{-1.8mm}The rapid evolution of wireless communication technologies and the advances towards 6G networks aim to satisfy the increasing data rate demands of emerging applications. Among these advances, millimeter-wave (mm-wave) and Terahertz (THz) frequency bands are key since they can provide abundant spectral resources for data transmission \cite{rappaport2019wireless}. However, communication in higher frequency bands usually suffers from significant propagation losses. As a result, massive multiple-input multiple-output (M-MIMO) antenna arrays have been used to boost signal power and minimize interference. Beam management, the process of assigning beams to users, is essential for optimizing network performance \cite{elsayed2020radio}.

Conventional position-based beam prediction methods face significant challenges due to user mobility and signal changes \cite{tian2023multimodal}. One solution is to integrate visual information into beam management to mitigate localization errors \cite{yao2022joint}. With the availability of data in more modalities, such as radar and light detection and ranging (LiDAR) data, there is a potential to further improve the accuracy of beam prediction by using more information from the user. On the other hand, multi-modal transformers (MMTs) have recently emerged, known for their remarkable ability to manage diverse data modalities \cite{xu2023multimodal}\cite{ahmad2023vision}. They facilitate information processing and feature extraction, which enhances contextual understanding and enables more accurate beam predictions \cite{charan2022vision}. 

Considering these benefits, we aim to leverage MMT in our work to effectively manage the diverse modalities in wireless communication systems. Nevertheless, adopting MMT poses challenges, particularly when increasing the number of labels or ground truth data for supervised learning, which can lead to lower accuracy. To address these challenges, we propose a beam management framework by integrating MMT with reinforcement learning (RL) in a two-step approach. The main contributions of this work are concluded as follows:
\begin{itemize}
\item In this work, we integrated MMT and RL as a two-step framework for multi-modal beam management. In the first step, we reduce the set of decisions by grouping the beam indices and then use MMT to predict the optimal group for transmission based on a variety of data modalities. Our second step is to use RL for beam decisions within each group to maximize throughput. This solves the existing challenge of effectively handling multi-modal data in beam management tasks. To the best of our knowledge, this is the first proposal to integrate MMT and RL to enhance beam management.
\item Unlike previous works focusing solely on beam prediction accuracy, this work also shows how efficient beam management can further enhance system throughput. The framework is tested with a real-world 6G dataset, demonstrating its practical effectiveness.
\end{itemize}

\vspace{-2mm}
\section{System Model and Problem Formulation}
\setlength{\textfloatsep}{10pt plus 1.0pt minus 2.0pt}
\begin{figure*}
\setlength{\belowcaptionskip}{-12pt}
\centering
\begin{minipage}{.2\textwidth}
  \centering
  \includegraphics[width=1.\linewidth]{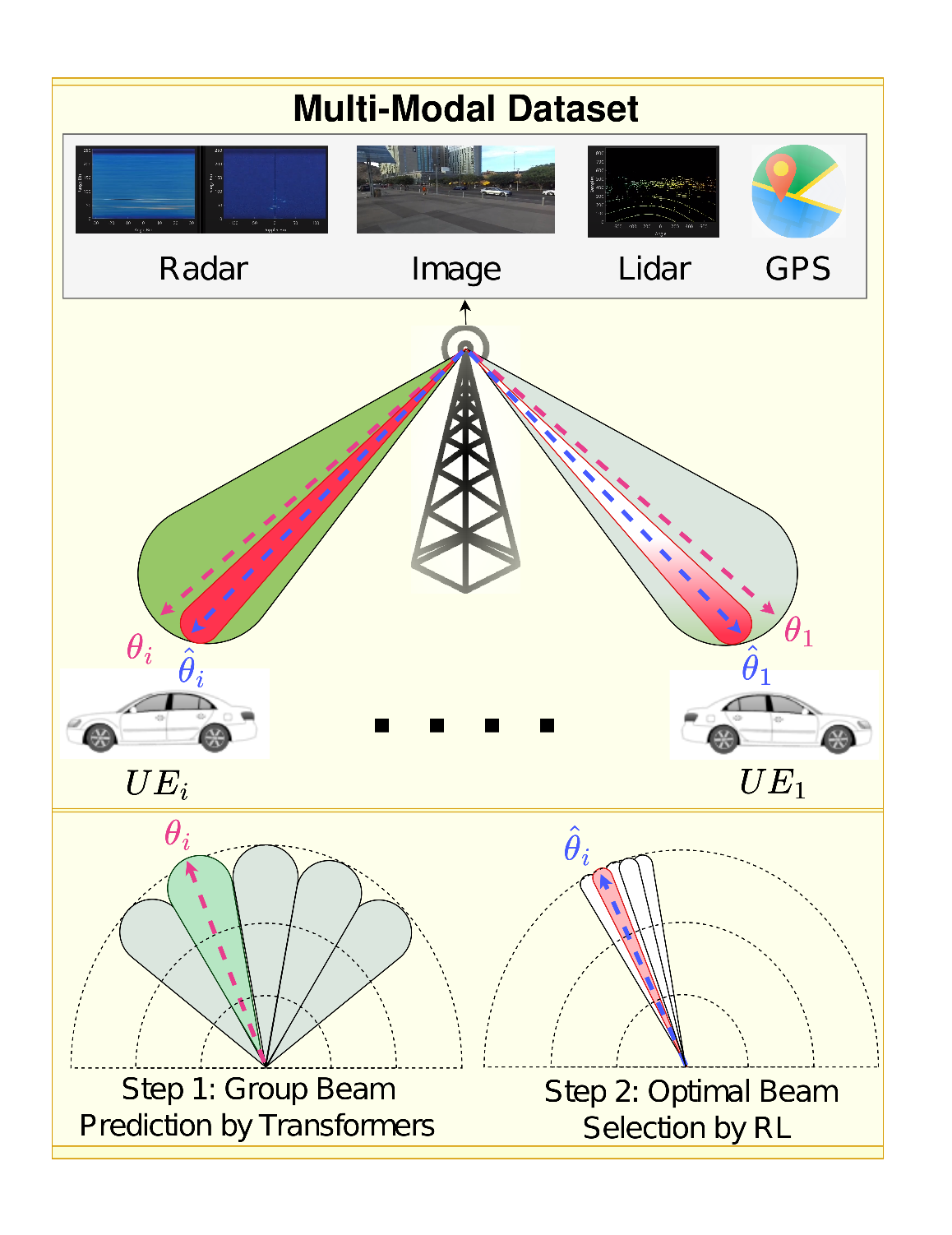}
  \captionof{figure}{A multimodal dataset and the use of radar, images, LiDAR, and GPS data for UE position determination.}
  \label{fig:test1}
\end{minipage}%
\begin{minipage}{.8\textwidth}
  \centering
  \includegraphics[width=.98\linewidth]{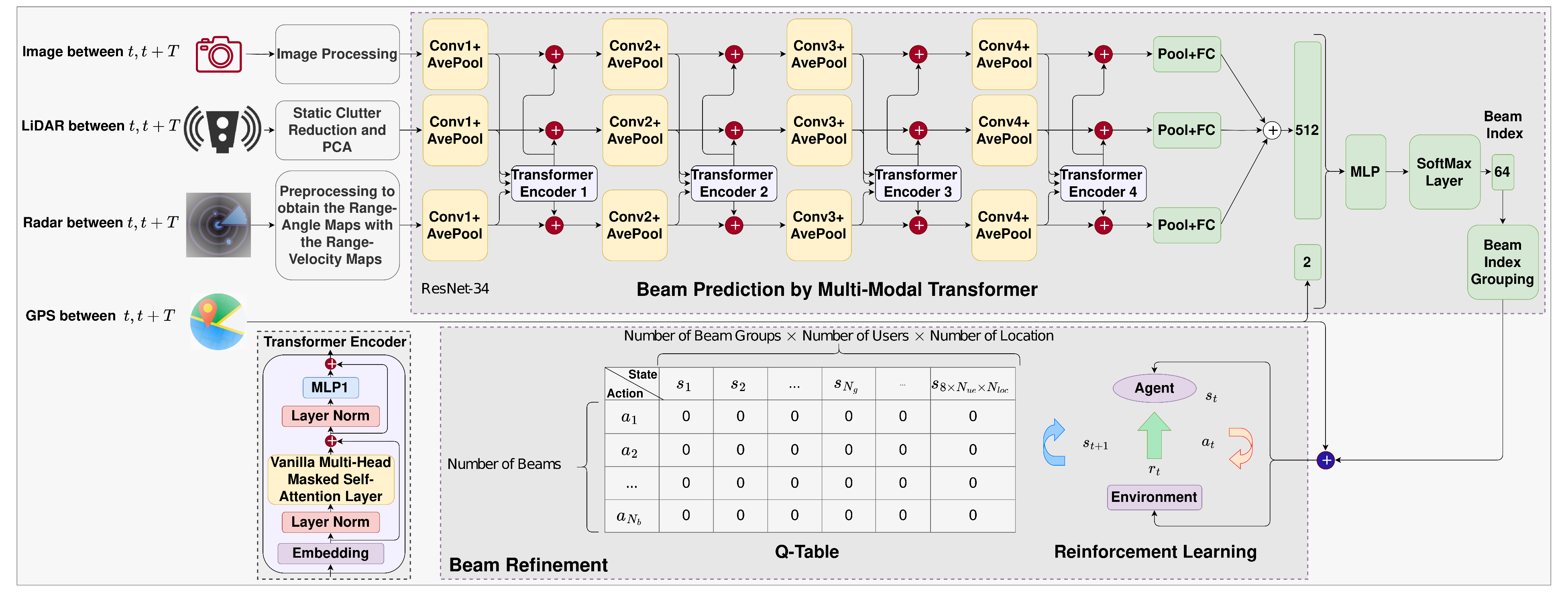}
  \captionof{figure}{Beam prediction and refinement framework using MMT and RL for optimal beam index selection.}
  \label{fig:test2}
\end{minipage}
\end{figure*}
\vspace{-1.8mm}\figurename{\ref{fig:test1}} illustrates our proposed two-step beam management framework. In this system, we assume that there is one base station (BS) and multiple user equipments (UEs). The framework aims to optimize the communication performance between the BS and UEs and maximize the throughput by selecting the optimal beam index.
We consider multi-modal data as the input of the framework, which includes vision data, radar data, LiDAR data, and GPS data.

In the first step, we categorize the available beam indices into multiple groups and use MMT to process various data modalities to predict the optimal beam group. In the second step, we utilize RL for beam decision-making within each group to maximize throughput. Through this framework, the advantages of MMT in handling multi-modal information and the advantages of RL in adapting to dynamic environments can be jointly utilized. In the following part of this section, the communication model for beam management is introduced and the problem formulation is described. 
In this work, a downlink orthogonal frequency-division multiplexing (OFDM) cellular system is considered. The received signal-to-interference-plus-noise ratio (SINR) at the $u^{th}$ UE can be calculated \cite{zhou2022knowledge}:
\vspace{-2mm}
\begin{equation}
SINR_u(\theta) = \frac{P_{u}(\theta)}{W_{u} . N_0 + \sum_{u^{'}\in U_{-u}}  P_{u^{'}}(\theta)},
\vspace{-2mm}
\end{equation}
where $P_u(\theta)$ denotes the received power of UE $u$, $W_u$ denotes the bandwidth, $N_0$ denotes the noise power density, $\theta$ denotes the angle between the UE and the BS, $U_{-u}$ denotes the set of UEs in a cell served by a BS except $u^{th}$ UE, and $P_{u^{'}}(\theta)$ denotes interference power from $U_{-u}$ users. The received power is \cite{balanis2016antenna}:
\vspace{-2mm}
\begin{equation} \label{eq:2}
P_u(\theta) = P_t \cdot \frac{G_t(\theta) \cdot G_r(\theta) \cdot \lambda^2}{(4\pi R)^2} \cdot PL,
\vspace{-2mm}
\end{equation}
where $P_{t}$ denotes the transmit power, $G_t$ denotes of gain of the transmitting antenna, $G_r$ denotes of gain of the receiving antenna, $\lambda$ denotes signal wavelength, $PL$ denotes path loss, and $R$ denotes distance between user and BS. $G_t$ is influenced by the directivity of the antenna array, also known as the array gain pattern, which can be defined as:
\vspace{-2mm}
\begin{equation} \label{eq:3}
U(\theta) = \left[ \frac {\sin(\frac{N_{a}}{2} k_{a}d_{a}\cos \theta)}{\frac{N_{a}}{2} k_{a}d_{a}\cos \theta} \right]^2,
\vspace{-2mm}
\end{equation}
where $N_{a}$ denotes the number of antenna elements, $k_{a} = \frac{2\pi}{\lambda}$ denotes the wavenumber, and $d_{a}$ denotes the spacing between the antennas \cite{balanis2016antenna}. With this equation, the gain (or directivity) of an antenna array can be described as a function of the beam angle $\theta$. Therefore, by selecting the optimal beam that steers the signal towards the desired UE and away from interfering users, we can increase $P_u$ at the desired UE. 

The system aims to dynamically adjust beamforming parameters to maximize SINR, thereby improving throughput which is described as:
\vspace{-2mm}
\begin{equation} \label{eq:4}
b(\theta)  = \sum_{u \in U} W_u  \log_2  ( 1 + SINR_u(\theta)).
\vspace{-2mm}
\end{equation}

Therefore, the corresponding throughput maximization problem can be formulated as:
\vspace{-2mm}
\begin{subequations} \label{eq:5}
\begin{eqnarray}
 \underset{\theta}{\mathrm{maximize}}  \sum_{i = 0}^{K_d} b(\theta)^{(i)} \label{subeq1}\\
 \textrm{subject to} \;\;\; (1)-(4) \\
 \sum_{n \in N_{ue}} x_{u, n} = 1 \\
 \theta = g(x_{u,n})
 \label{subeq2}
 \vspace{-2mm}
\end{eqnarray}
\end{subequations}
where $K_d$ denotes the number of iterations, $N_{ue}$ denotes the number of users, and $x_{u, n}$ is a binary variable, with $x_{u, n}=1$  indicating that the beam antenna index $n$ is allocated to UE $u$; otherwise, $x_{u, n}$ = 0. Also, $g(x_{u,n})$ is a look-up table mapping $x_{u,n}$  to $\theta$. This ensures that each UE is associated with a specific beam antenna index, governed by the values of $g(x_{u,n})$. Eq. (\ref{subeq2})  guarantees one beam can only be allocated to the UE.

\vspace{-2mm}
\section{MMT and RL-based Beam Management \label{section4}}
\vspace{-1.8mm}In this section, we present the detailed implementation of our proposed method, as depicted in \figurename{\ref{fig:test2}}. 
\vspace{-2mm}
\subsection{Group Beam Prediction by MMT}
In the first step, as illustrated in \figurename{\ref{fig:test2}}, the 64 optional beam angles are divided into $N_{g} = 8$ groups, and a pre-trained MMT proposed in \cite{tian2023multimodal} is utilized for beam group prediction. The MMT is firstly fine-tuned through supervised learning to choose the group based on similar predicted features ($\theta_i$). The input of the MMT is a multi-modal data source, including images, LiDAR data, and radar data. The output is defined as the group beam indices. The architecture of the MMT model includes two parts: the convolutional neural networks (CNNs) and the transformer models. CNNs are used for feature extraction, while transformers are used for learning the relationships between these features. This combination allows the model to capture complex patterns in multi-modal data and facilitates effective beam clustering. Additionally, this architecture also includes a deep residual network (ResNet) for encoding features. The fusion of feature maps from different modalities creates a feature vector. Moreover, we concatenate calibrated GPS locations (longitude and latitude) with the vector and pass them through multi-layer perception (MLP) layers to produce weights for 64 beam indices using the softmax function \cite{tian2023multimodal}.  This step efficiently reduces the search space for the subsequent RL step. 

With self-attention mechanisms, transformers are well-suited for capturing correlations and relationships in data from dynamic environments. In the MMT model, encoders process different modalities before passing them to the transformer, allowing the system to handle multiple sources of information simultaneously. The design of MMTs allows for scalability, as additional encoders can be easily added to handle new modalities due to their transformer-based architecture. The MMT architecture includes a ResNet32 backbone, pre-trained on the ImageNet dataset, providing a robust foundation for understanding multi-modal data. Fine-tuning on the DeepSense 6G dataset further adapts the model to dynamic wireless communication environments, enhancing its ability to perform tasks like beam management in challenging scenarios.

The flexibility of MMTs is also demonstrated by their performance with varying user numbers in a complex 6G environment during the simulations. Moreover, real-world datasets may have challenges such as varying data quality and synchronization issues. MMTs leverage self-attention to align key information across modalities and compensate for these issues.

\vspace{-2mm}
\subsection{Optimal Beam Selection using RL}
In the second step, the RL agent operates in a simulated environment based on a real-world scenario, aiming to select the optimal beam among $N_{b} = 8$ beams within each group ($\hat{\theta}_i$), where $N_{b}$ denotes the number of beams within each group. By incorporating inputs from the MMT model and GPS data, the agent learns an optimal policy for beam selection through iterative interactions with the real-world environment.
The agent observes the state, makes decisions by selecting the optimal beam, receives the reward, and updates its policy iteratively. This feedback mechanism allows the agent to refine its decision-making policy over time to maximize the reward function.
In this work, we employ Q-learning, where the RL agent uses a Q-table to store rewards. This choice is driven by our finite action and state spaces, which are well-suited for our approach. The action space consists of 8 options, and the state space is $\mathcal{O}(N_{b}\times N_{g}\times N_{ue} \times N_{loc})$, where $N_{g}$ denotes the number of groups and $N_{loc}$ denotes the number of location samples. The state space depends on the number of users. This setup allows us to represent all possible state-action pairs without requiring function approximation, making it more appropriate than Deep Reinforcement Learning (DRL). To avoid adding further complexity to the MMT model, we chose Q-learning over DRL. Q-learning offers significant computational efficiency that allows us to maintain high beam management accuracy while minimizing computational complexity without the need for intensive training processes associated with DRL. The Markov decision process (MDP) is defined as follows:

\begin{itemize}
\item \textbf{State:} The state can be represented as $s_{t} = (u_{t}, v_{t})$, where: $u_{t}$ denotes the locations of the users at time $t$, and $v_{t}$ is a set of group indices derived from the output of the MMT at time $t$.
\item \textbf{Action:} The action of the agent is to decide which beam to assign to a user during each time slot. It can be represented as $a_{t} = b_{t}$, where $b_{t}$ denotes the beam index that the agent chooses to allocate to a user at time $t$.
\item \textbf{Reward:} 
If the system throughput exceeds a set threshold ($b > r_{th}$), the agent receives a constant reward; otherwise, it incurs a penalty. This threshold-based reward approach helps prevent issues like reward sparsity or misleading signals, ensuring a more stable learning environment for the agent\cite{wang2019reinforcement}.
\end{itemize}
At each time step $t$, the agent observes the environment's state $s_{t}$ and selects an action $a_{t}$, leading to a transition to a new state $s_{{t}+1}$ and receiving a reward $r_{t}$. The primary objective in RL is to optimize the cumulative reward $C_{t}$ over time, defined as $C_{t} = \sum_{k=0}^{\infty} \gamma^k r_{k+t}$, where $\gamma$ ($\gamma \in [0,1]$) is the discount factor for future rewards.

The action-value function $Q_{\pi}(s, a)$ represents the expected return when taking action $a$ in state $s$ following policy $\pi$. RL seeks to determine the optimal policy $\pi^*$ that maximizes the $Q$-function across all possible policies. In Q-learning, the Q-values are updated by $Q(s_{{t}},a_{{t}}) \leftarrow Q(s_{t},a_{t}) +  \alpha [r_{t+1} + \gamma \max_{a} Q(s_{t+1},a) - Q(s_{t},a_{t})]$ where $\alpha$ is the learning rate and $r_{t+1}$ is the reward at time $t+1$. Accurate estimation of the $Q$-function leads to determining the optimal policy $\pi^*$ at a given state $s$, selecting the action $a$ with the highest value.

\vspace{-2mm}
\subsection{DeepSense 6G Dataset and Data Preprocessing}
The DeepSense 6G dataset \cite{charan2022multi} encompasses multi-modal data across diverse deployment scenarios. In this work, we use the Scenario 32 dataset proposed in \cite{alkhateeb2023deepsense}, collected from a two-way city street environment with vehicles and pedestrians, captured by two synchronized data collection units positioned on opposite sides of the street. This scenario represents real-world urban wireless communication challenges and is available online.

Before feeding the multi-modal data into the beam management framework, preprocessing ensures data quality and consistency through cleaning, normalization, and feature scaling. The dataset includes RGB images with segmented masks indicating object locations, and LiDAR data capturing 3D point clouds that describe the environment and obstacles from various angular views. 
Following previous work \cite{dantas2023split}, we apply Static Clutter Reduction (SCR) \cite{shin2020multiple} to the LiDAR data to remove stationary and irrelevant objects, reducing noise and improving data quality. The filtered point cloud retains only key spatial measurements, after which Principal Component Analysis (PCA) is used to extract the $l$ most significant features, reducing dimensionality while preserving essential information.
For radar data, the goal is to determine the distance, angles, and velocity of moving objects. To extract explicit velocity information, we concatenate the Range-Angle and Range-Velocity Maps, preserving speed data for moving cars. Radar signals reliably measure speed regardless of weather or lighting conditions.

\vspace{-2mm}
\section{Simulation and Results}
\vspace{-1.8mm}
\subsection{Parameter Settings}
In the simulation, the transmit signal power is set to $P_{t}$ = 40 dBm, and the background noise power is set to $N_{0}$ = -10 dBm. We consider a single BS equipped with $N_{a}=64$. By default, we consider $N_{ue} = 5$ and $N_{loc} = 5$. However, for calculating throughput, we evaluate scenarios with $N_{ue} = 10, \,15, \,20,$  and $25$ UEs. The RL agent employs an epsilon-greedy policy to balance exploration and exploitation. Initially, a high epsilon value of 0.9 promotes the exploration of various beam configurations. As episodes progress, the epsilon value decreases to 0.6, prioritizing the use of the best-known configurations based on prior learning. For simulation purposes, we use GPS information to determine the distance between the BS and users. The channel model 128.1 + 37.6 log(distance(km)) represents a Line-of-Sight (LOS) path loss model used in wireless communication systems. The number of beam measurements $k$ refers to the minimum set of beams that must be evaluated to accurately determine the optimal beam, balancing between minimizing overhead and maximizing selection accuracy; in our simulation, we choose $k$ from 1 to 5. Also, we choose $l=14$.
For MMT, we partition datasets into a 90\% training set and a 10\% validation set for optimizing model weights. The validation set is used for hyperparameter configurations. We select epochs = 150, learning rate = 0.0005, and batch size = 6. For RL, we use a discount rate of 0.9 and a learning rate of 0.001 \cite{tian2023multimodal}. 

There is a difference in time scales between the MMT and RL's processing interval. The MMT processes incoming data from the environment every $\textit{\textup{T}} = 100$ ms and predicts beam groups. The RL updates every four TTIs, which is approximately 0.5716 ms. Therefore, the RL episode is $K_d = \frac{100}{0.5716} \approx 174$ steps. It allows the MMT to process larger windows of time while the RL agent operates at shorter and more frequent intervals. The results are averaged over 10 Monte Carlo rounds of training.

\vspace{-2mm}
\subsection{Simulation Results}


\begin{figure*}\setlength{\belowcaptionskip}{-6pt}
  \begin{subfigure}{0.3\textwidth}
    \includegraphics[width=\linewidth]{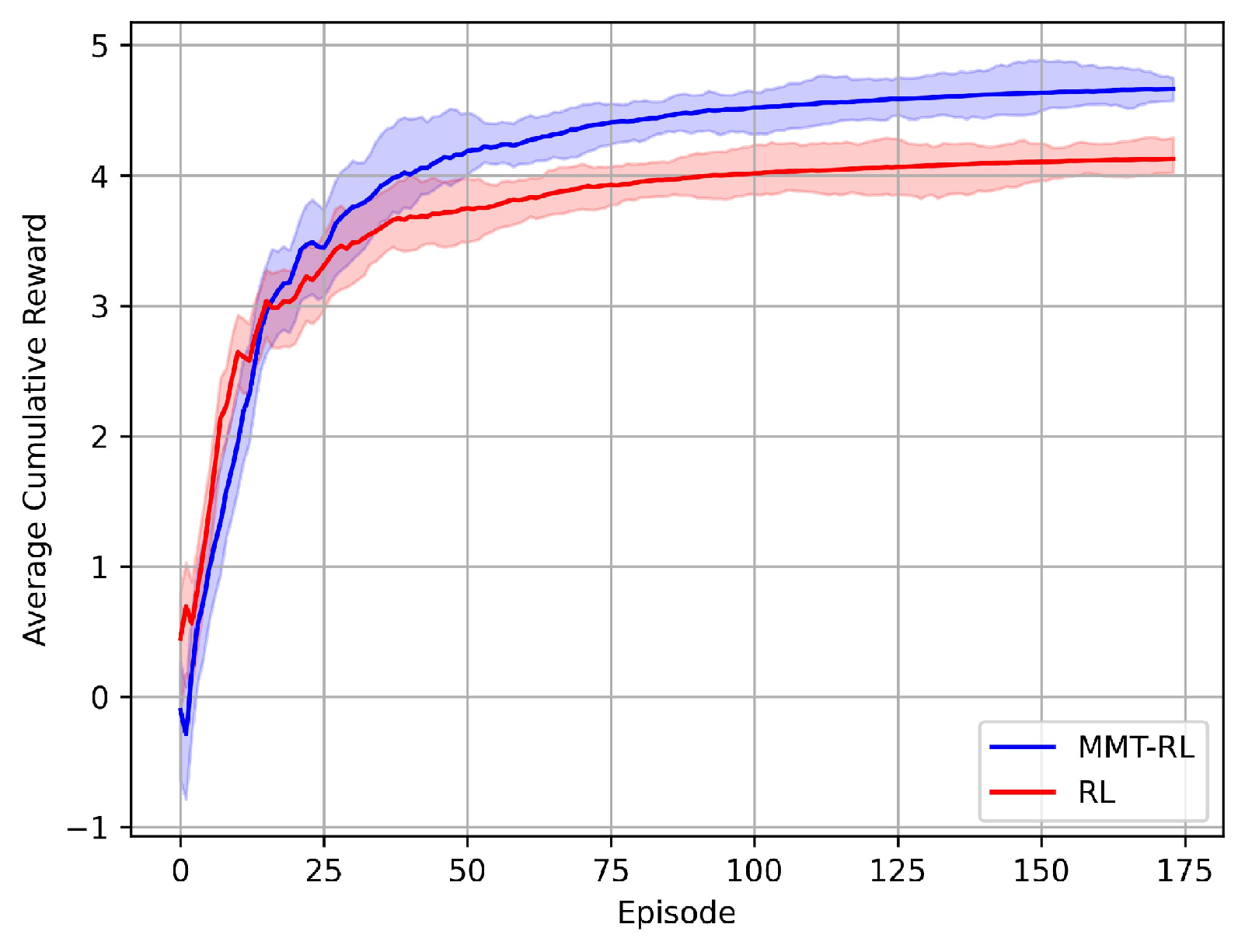}
    \caption{The average cumulative reward.} \label{fig:Reward}
  \end{subfigure}%
  \hspace*{\fill}   
  \begin{subfigure}{0.3\textwidth}
    \includegraphics[width=\linewidth]{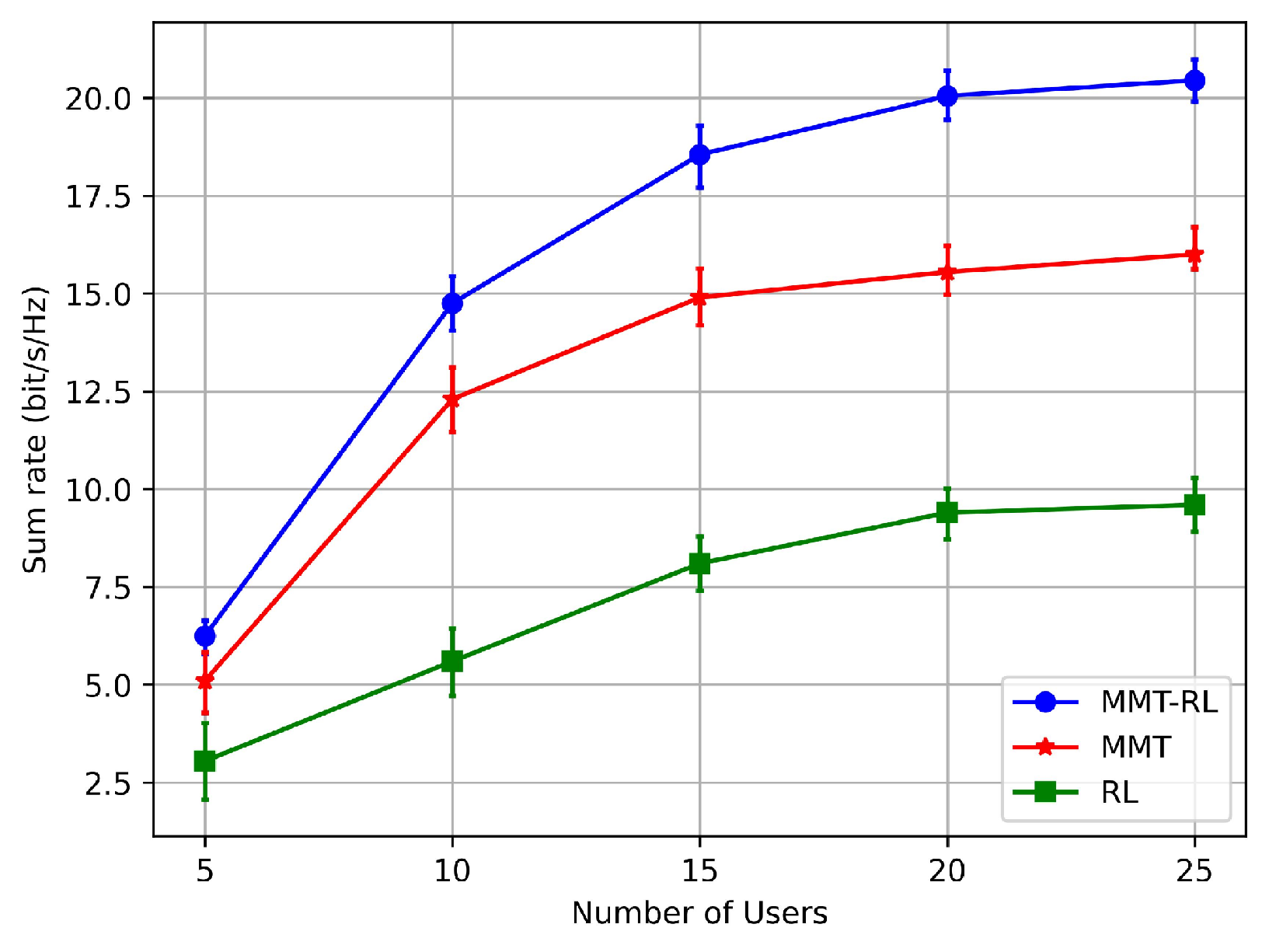}
    \caption{Throughput.} \label{fig:ThroughputPerUser}
  \end{subfigure}%
  \hspace*{\fill}   
  \begin{subfigure}{0.3\textwidth}
    \includegraphics[width=\linewidth]{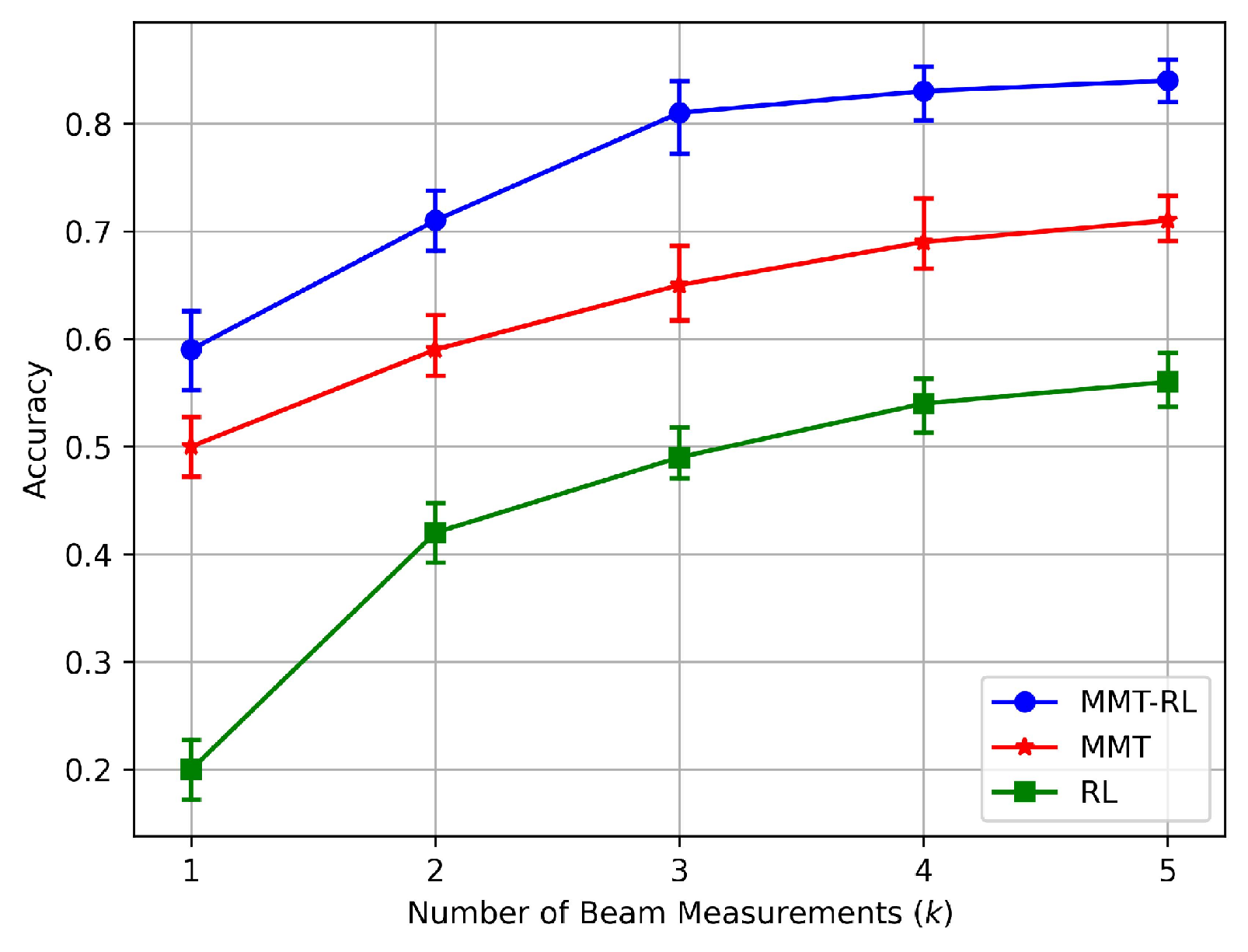}
    \caption{Beam selection accuracy} \label{fig:AccuracyPerUser}
  \end{subfigure}
\captionsetup{belowskip=-20pt}
\caption{Simulation results comparison.} \label{fig:1}
\end{figure*}

\figurename{\ref{fig:Reward}} compares the convergence performance of our proposed MMT-RL method with a baseline RL-only approach when the number of users is 5. The x-axis represents the episode number in the simulation, and the y-axis represents the average cumulative reward. In the RL-based baseline, the agent selects from 64 optional beams directly, whereas the MMT-RL approach benefits from a reduced search space by initially predicting beam groups, narrowing down to 8 beams per group. This significant reduction allows the RL agent to focus on a more targeted set of beam candidates, leading to a higher average cumulative reward compared to the baseline. The MMT-RL approach demonstrates greater effectiveness in maximizing reward due to its efficient beam management strategy.
The average cumulative reward plot further illustrates the agent's performance improvement over time. As shown in \figurename{\ref{fig:Reward}}, both methods exhibit a sharp increase in cumulative reward during the initial episodes, indicating exploration of different strategies. The curves stabilize after episode 50, indicating that the agent has successfully learned and is now focused more on exploiting rather than exploring new ones.

\figurename{\ref{fig:ThroughputPerUser}} compares the performance of the throughput of MMT-RL with baseline MMT and RL-based approaches with varying numbers of users. The x-axis represents the number of users in the simulation, and the y-axis denotes the achieved throughput, indicated by spectral efficiency (SE). We assess beam management performance in both the baseline MMT and RL-based methodologies using 64 narrow beam indices and calculate SE. As evident from \figurename{\ref{fig:ThroughputPerUser}}, the MMT-RL approach achieves significantly higher throughput compared to the baseline MMT and RL approaches. This indicates that our proposed method is more effective at maximizing throughput in the simulated wireless communication network. Even without RL, the use of MMT in both methods might provide some throughput advantage over traditional beam management techniques. The reason for this is that MMT can capture complex relationships between different modalities of data that can influence beam management. Our proposed method demonstrates superior throughput in wireless network scenarios compared to the MMT and RL methods. With 5 users, our method achieves 6.25 bit/s/Hz, while MMT and RL achieve 5.1 bit/s/Hz and 3.05 bit/s/Hz, respectively. When the number of users increases to 25, our method achieves 20.45 bit/s/Hz, compared to MMT's 16.3 bit/s/Hz and RL's 9.6 bit/s/Hz. These results highlight our method's effectiveness in maintaining higher throughput as user density increases.




\figurename{\ref{fig:AccuracyPerUser}} compares the beam selection accuracy of MMT-RL with baseline MMT and RL-based approaches against $k$. In the MMT approach, supervised learning was employed without considering throughput optimization. Despite RL's inability to utilize multi-modal information and reliance solely on GPS, MMT-RL leverages the strengths of both MMT and RL to enhance accuracy and throughput. Increasing the value of $k$ improves accuracy across all models by exploring more potential beams, reducing the chance of missing optimal selections. However, this also increases overhead costs. Our proposed method achieves top-1, top-3, and top-5 beam selection accuracy of 59.5\%, 81.0\%, and 84.2\%, respectively. By choosing $k$=5, our approach reaches 84.2\% accuracy, which is 13\% higher than MMT and 28\% higher than RL.

\vspace{-1.8mm}
\subsection{Computational Complexity Analyses}
In this section, we analyze the computational complexity of the proposed method based on runtime analysis. We utilized an NVIDIA RTX 3060 GPU and a CPU featuring Intel Core i7 11370H for our experiments.
The inference time of the proposed method is shown in Table \ref{table:inferencetime2}. The results indicate that the single inference time for the MMT model is 74.8 ms, while the RL inference time is 0.031 ms for 5 users and 0.155 ms for 25 users. The combined inference time for the proposed method ranges from 74.831 to 74.955 ms, meaning the RL step adds less than 0.2\% to the overall processing time. This reinforces the point that while there is an additional step in the beam management process in our proposed method, its contribution to the overall runtime is negligible.
\begin{table}[h]
\centering
\renewcommand\thetable{I}
\caption{Inference Time}
\label{table:inferencetime2}
\begin{tabular}{|c|c|c|c|}
\hline
\textbf{Method}  & \textbf{Inference Time (milliseconds)}   \\ \hline
MMT             & 74.8                \\ \hline
RL              & 0.031 $\sim$  0.155           \\ \hline
Proposed Method   & 74.831 $\sim$  74.955          \\ \hline
\end{tabular}
\end{table}

\vspace{-2mm}
Potential optimizations for future work include model pruning to reduce MMT parameters and quantization to lower memory usage and improve inference speed.

\vspace{-2mm}
\section{Conclusion}
\vspace{-1.8mm}
Our study presented a novel approach integrating MMT and RL for multi-modal beam management in wireless communication systems. By incorporating RL, we addressed the limitation of previous MMT-based approaches by reducing the decision space. Our approach combines MMT's feature extraction with RL's adaptability to dynamic environments, offering a promising solution for beam management. The results were obtained using the Deepsense 6G dataset, which provides real-world data for robust evaluation. Through comparisons with baseline methods, we demonstrate the superiority of the proposed method, achieving higher beam selection accuracy and throughput performance, contributing to the development of more efficient wireless communication systems.

\vspace{-2mm}

\section*{Acknowledgment}
\vspace{-2mm}
This work has been supported by NSERC Canada Research Chairs program, MITACS and Ericsson.
\vspace{-2mm}

\bibliographystyle{IEEEtran}
\bibliography{bibliography.bib}

\begin{thebibliography}{10}
\providecommand{\url}[1]{#1}
\csname url@samestyle\endcsname
\providecommand{\newblock}{\relax}
\providecommand{\bibinfo}[2]{#2}
\providecommand{\BIBentrySTDinterwordspacing}{\spaceskip=0pt\relax}
\providecommand{\BIBentryALTinterwordstretchfactor}{4}
\providecommand{\BIBentryALTinterwordspacing}{\spaceskip=\fontdimen2\font plus
\BIBentryALTinterwordstretchfactor\fontdimen3\font minus \fontdimen4\font\relax}
\providecommand{\BIBforeignlanguage}[2]{{%
\expandafter\ifx\csname l@#1\endcsname\relax
\typeout{** WARNING: IEEEtran.bst: No hyphenation pattern has been}%
\typeout{** loaded for the language `#1'. Using the pattern for}%
\typeout{** the default language instead.}%
\else
\language=\csname l@#1\endcsname
\fi
#2}}
\providecommand{\BIBdecl}{\relax}
\BIBdecl

\bibitem{rappaport2019wireless}
T.~S. Rappaport, Y.~Xing, O.~Kanhere, S.~Ju, A.~Madanayake, S.~Mandal, A.~Alkhateeb, and G.~C. Trichopoulos, ``Wireless communications and applications above 100 ghz: Opportunities and challenges for 6g and beyond,'' \emph{IEEE access}, vol.~7, pp. 78\,729--78\,757, 2019.

\bibitem{elsayed2020radio}
M.~Elsayed and M.~Erol-Kantarci, ``Radio resource and beam management in 5g mmwave using clustering and deep reinforcement learning,'' \emph{GLOBECOM 2020-2020 IEEE Global Communications Conference}, pp. 1--6, 2020.

\bibitem{tian2023multimodal}
Y.~Tian, Q.~Zhao, F.~Boukhalfa, K.~Wu, F.~Bader \emph{et~al.}, ``Multimodal transformers for wireless communications: A case study in beam prediction,'' \emph{arXiv preprint arXiv:2309.11811}, 2023.

\bibitem{yao2022joint}
Y.~Yao, H.~Zhou, and M.~Erol-Kantarci, ``Joint sensing and communications for deep reinforcement learning-based beam management in 6g,'' \emph{GLOBECOM 2022-2022 IEEE Global Communications Conference}, pp. 5019--5024, 2022.

\bibitem{xu2023multimodal}
P.~Xu, X.~Zhu, and D.~A. Clifton, ``Multimodal learning with transformers: A survey,'' \emph{IEEE Transactions on Pattern Analysis and Machine Intelligence}, 2023.

\bibitem{ahmad2023vision}
I.~Ahmad, A.~R. Khan, R.~N.~B. Rais, A.~Zoha, M.~A. Imran, and S.~Hussain, ``Vision-assisted beam prediction for real world 6g drone communication,'' \emph{2023 IEEE 34th Annual International Symposium on Personal, Indoor and Mobile Radio Communications (PIMRC)}, pp. 1--7, 2023.

\bibitem{charan2022vision}
G.~Charan, T.~Osman, A.~Hredzak, N.~Thawdar, and A.~Alkhateeb, ``Vision-position multi-modal beam prediction using real millimeter wave datasets,'' \emph{2022 IEEE Wireless Communications and Networking Conference (WCNC)}, pp. 2727--2731, 2022.

\bibitem{zhou2022knowledge}
H.~Zhou and M.~Erol-Kantarci, ``Knowledge transfer based radio and computation resource allocation for 5g ran slicing,'' \emph{2022 IEEE 19th Annual Consumer Communications \& Networking Conference (CCNC)}, pp. 617--623, 2022.

\bibitem{balanis2016antenna}
C.~A. Balanis, \emph{Antenna theory: analysis and design}.\hskip 1em plus 0.5em minus 0.4em\relax John wiley \& sons, 2016.

\bibitem{wang2019reinforcement}
R.~Wang, O.~Onireti, L.~Zhang, M.~A. Imran, G.~Ren, J.~Qiu, and T.~Tian, ``Reinforcement learning method for beam management in millimeter-wave networks,'' \emph{2019 UK/China Emerging Technologies (UCET)}, pp. 1--4, 2019.

\bibitem{charan2022multi}
G.~Charan, U.~Demirhan, J.~Morais, A.~Behboodi, H.~Pezeshki, and A.~Alkhateeb, ``Multi-modal beam prediction challenge 2022: Towards generalization,'' \emph{arXiv preprint arXiv:2209.07519}, 2022.

\bibitem{alkhateeb2023deepsense}
A.~Alkhateeb, G.~Charan, T.~Osman, A.~Hredzak, J.~Morais, U.~Demirhan, and N.~Srinivas, ``Deepsense 6g: A large-scale real-world multi-modal sensing and communication dataset,'' \emph{IEEE Communications Magazine}, pp. 122--128, 2023.

\bibitem{dantas2023split}
Y.~Dantas, P.~E. Iturria-Rivera, H.~Zhou, Y.~Ozcan, M.~Bavand, M.~Elsayed, R.~Gaigalas, and M.~Erol-Kantarci, ``Split learning for sensing-aided single and multi-level beam selection in multi-vendor ran,'' \emph{GLOBECOM 2023-2023 IEEE Global Communications Conference}, pp. 6652--6657, 2023.

\bibitem{shin2020multiple}
M.~Shin and H.~Son, ``Multiple sensor linear multi-target integrated probabilistic data association for ultra-wide band radar,'' \emph{IEEE Access}, vol.~8, pp. 227\,161--227\,171, 2020.

\end{thebibliography}

\end{document}